# One-Dimensional Moiré Excitons in Transition-Metal Dichalcogenide Heterobilayers

Yusong Bai,[1,*] Lin Zhou,[1,2,*] Jue Wang,[1] Wenjing Wu,[1] Leo McGilly,[3] Dorri Halbertal,[3] Chiu Fan B. Lo [3], Fang Liu,[1] Jenny Ardelean,[4] Pasqual Rivera,[5] Nathan R. Finney,[4] Xuchen Yang,[6] Dmitri N. Basov,[3] Wang Yao,[6] Xiaodong Xu,[5] James Hone,[4] Abhay Pasupathy,[3,†] Xiaoyang Zhu[1,†]

[1] Department of Chemistry, Columbia University, New York, NY 10027, USA
[2] College of Engineering and Applied Sciences, Nanjing University, Nanjing 210093, China
[3] Department of Physics, Columbia University, New York, NY 10027, USA
[4] Department of Mechanical Engineering, Columbia University, New York, NY 10027, USA
[5] Department of Physics and Department of Materials Science and Engineering, University of Washington, Seattle, WA 98195, USA
[6] Department of Physics, University of Hong Kong, Hong Kong, China

* These authors contributed equally.

† To whom correspondence should be addressed. xyzhu@columbia.edu, apn2108@columbia.edu

**The formation of interfacial moiré patterns from angular and/or lattice mismatch has become a powerful approach to engineer a range of quantum phenomena in van der Waals heterostructures[1–4]. For long-lived and valley-polarized interlayer excitons in transition-metal dichalcogenide (TMDC) heterobilayers[5], signatures of quantum confinement by the moiré landscape have been reported in recent experimental studies[6–9]. Such moiré confinement has offered the exciting possibility to tailor new excitonic systems, such as ordered arrays of zero-dimensional (0D) quantum emitters[1] and their coupling into topological superlattices[1,10]. A remarkable nature of the moiré potential is its dramatic response to strain, where a small uniaxial strain can tune the array of quantum-dot-like 0D traps into parallel stripes of one-dimensional (1D) quantum wires[4]. Here, we present direct evidence for the 1D moiré potentials from real space imaging and the corresponding 1D moiré excitons from photoluminescence (PL) emission in MoSe$_2$/WSe$_2$ heterobilayers. Whereas the 0D moiré excitons display quantum emitter-like sharp PL peaks with circular polarization, the PL emission from 1D moiré excitons has linear polarization and two orders of magnitude higher intensity. The results presented here establish strain engineering as a powerful new method to tailor moiré potentials as well as their optical and electronic responses on demand.**

We choose the MoSe$_2$/WSe$_2$ heterobilayer as a model system because moiré interlayer excitons trapped in 0D moiré potential have recently been reported[8,9]. The heterobilayer is obtained from the transfer stacking method[11,12], either sandwiched between thin hexagonal boron nitride (h-BN) flakes (**Fig. 1a**), or with bottom h-BN but no top h-BN capping. TMDC heterobilayers are intrinsically endowed with moiré landscapes, as illustrated by the hexagonal moiré pattern in **Fig. 1b** with period $b \approx a/\sqrt{(\Delta\theta)^2 + \delta^2}$, where $a$ is the monolayer lattice constant and $\delta$ is the lattice mismatch and $\Delta\theta$ is the twist angle[1,13]. The latter can be determined by second harmonic generation (SHG)[14]. This moiré pattern can be strongly modified by applying differential strain (i.e. inequivalent strains on the two constituting layers). Specifically, a modest uniaxial strain can transform the moiré pattern into a hierarchy of a primary structure consisting of elongated ovals (arrows) that are parallel to each other to form a secondary structure of pseudo 1D stripes (dashed lines, **Fig. 1c**). For sufficiently small Δθ (from 0° or 60°), the differential and uniaxial tensile strain merges the elongated ovals into 1D stripes, **Fig. 1d**[4].

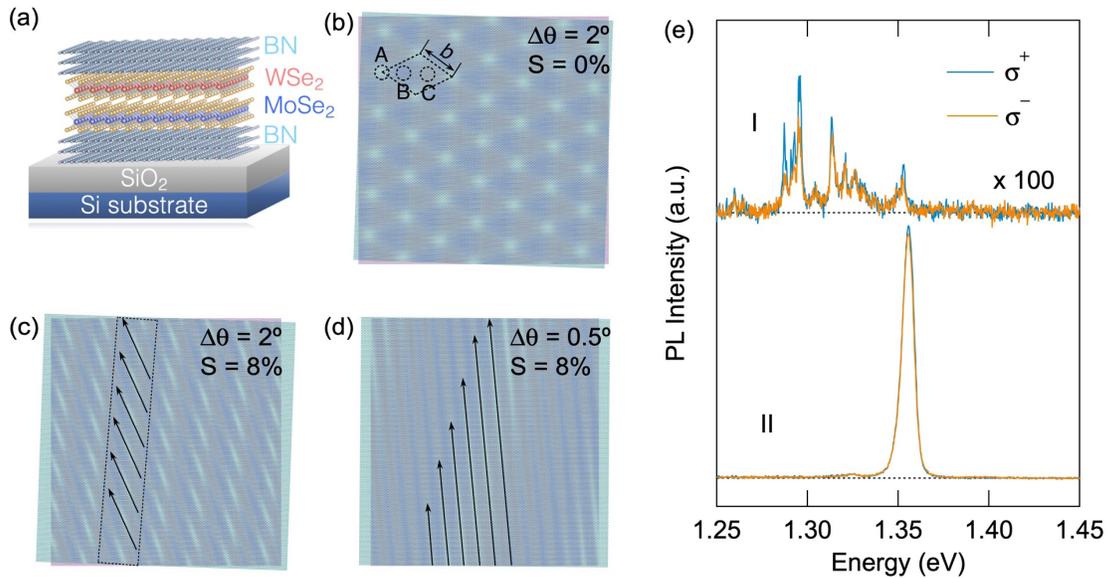

**Fig. 1 | Hexagonal vs quasi-1D moiré patterns and the associated photoluminescence spectra. (a)** Schematic of the vertically stacked WSe$_2$/MoSe$_2$ heterobilayer with BN encapsulation on SiO$_2$/Si substrate. Schematics of (**b**) hexagonal and (**c-d**) quasi-1D (strained) moiré superlattices with uniaxial strain (S) and with twist angles $\Delta\theta$ = 2° and 0.5°, respectively. The quasi-1D moiré in (c) or (d) is formed by a 8% uniaxial tensile strain on one (red) of the two monolayers. (**e**) Representative PL spectra from two WSe$_2$/MoSe$_2$ heterobilayer samples with h-BN encapsulation. Blue and orange represent $\sigma^+$ and $\sigma^-$ circularly polarized emission under $\sigma^+$ excitation. PL spectra were obtained with excitation density $n_{ex}$ ~ 6.1×10$^9$ cm$^{-2}$ at 4 K (h$\nu$ = 1.65 eV, repetition rate 76 MHz, pulse duration ~ 150 fs, same in all PL measurements).

As we establish below, the 0D and 1D moiré patterns give rise to two distinctive types of photoluminescence (PL) spectra at low excitation densities ($n_{ex} \leq 1\times10^{11}$ cm$^{-2}$). On ~20% of samples, we observe sharp PL peaks with the full-width-at-half-maximum (FWHM) ≤ 1 meV. This type-I features exhibit strong circular polarization under circularly polarized excitation (**Fig. 1e**, upper), in agreement with the recent report of Seyler et al.[8]. On the majority of samples (~80%), we observe a single broader PL peak with FWHM = 8±2 meV, labeled as type-II (**Fig. 1e**, lower), in agreement with all other previous reports on interlayer excitons in the WSe$_2$/MoSe$_2$ heterobilayer[12,15,16]. We find that in contrast to type-I PL emission, there is little circular but strong linear polarization from type-II. For either AA- or AB-stacked samples, the type-II PL intensity is over two-orders of magnitude higher than that of type-I.

To establish the origins of the two types of PL features, we directly image the moiré superlattice structure using piezoresponse force microscopy (PFM). TMDC monolayers and bilayers possess strong piezoelectric responses[17,18]. The moiré landscape in the heterobilayer modulates the piezoelectric tensor elements, leading to lateral deformations of the TMDC heterostructure in the presence of a vertically applied AC electric field. This electromechanical coupling is the basis for real space imaging by PFM of the moiré patterns typically not visible in conventional atomic force microscopy (AFM)[19]. **Figure 2a** presents PFM images of the BN/MoSe$_2$/WSe$_2$ stack, without top h-BN capping, on the soft polymer transfer tape before transferring onto the hard Si/SiO$_2$ substrate. We select five locations - white squares labeled 1-5 on the AFM image (**Fig. 2b**). On the soft polymer substrate, PFM imaging reveals hexagonal moiré patterns, as shown in **Fig. 2 c1-e1**, corresponding to pots 1-3. Additional PFM images of hexagonal moiré patterns from spots 4 and 5 on this sample and other samples are shown in **Figs. S2b &d** and **Figs. S3-a**, respectively. Fast Fourier transforms (FFTs, bottom right insets) of the images in **Fig. 2 c1-e1** identify the 6-fold symmetry. It reveals a superlattice constant of $b$ = 17.0±1.7 nm, corresponding to Δθ = 58.9±0.1° in good agreement with the AB stacking of Δθ = 59.4±0.8° determined in SHG (**Fig. S1-a**).

After the BN/MoSe$_2$/WSe$_2$ stack is transferred from the polymer tape to the Si/SiO$_2$ substrate, we carry out PFM imaging on exactly the same locations before transfer. While some of the locations (1&4) retain the hexagonal moiré pattern, as shown in **Fig. 2-c2** and **Fig. S2e**, other spots (2, 3, and 5) are transformed into distorted 1D moiré superlattices (see **Fig. 2-d2, e2**, **Fig. S2-c**, and **Fig. S3-b**). These 1D moiré patterns are consistent with the presence of uniaxial and differential strain, as illustrated in **Fig. 1c** and **1d**. From extensive PFM imaging of multiple samples, we estimate that ~20% of the locations retains the original hexagonal 0D moiré patterns and ~80% is transformed from 0D to 1D moiré patterns. In **Fig. 2-d2**, the formation of large 1D stripes with longitudinal periodicity of 49.0±22.0 nm is confirmed in the 2-fold symmetry in the FFT (inset). In **Fig. 2-e2**, PFM imaging resolves the hierarchical 1D moiré patterns consisting of elongated ovals that are parallel to each other (periodicity 9.3±0.6 nm) to form a secondary structure of 1D stripes (periodicity 25.6±5.6 nm).

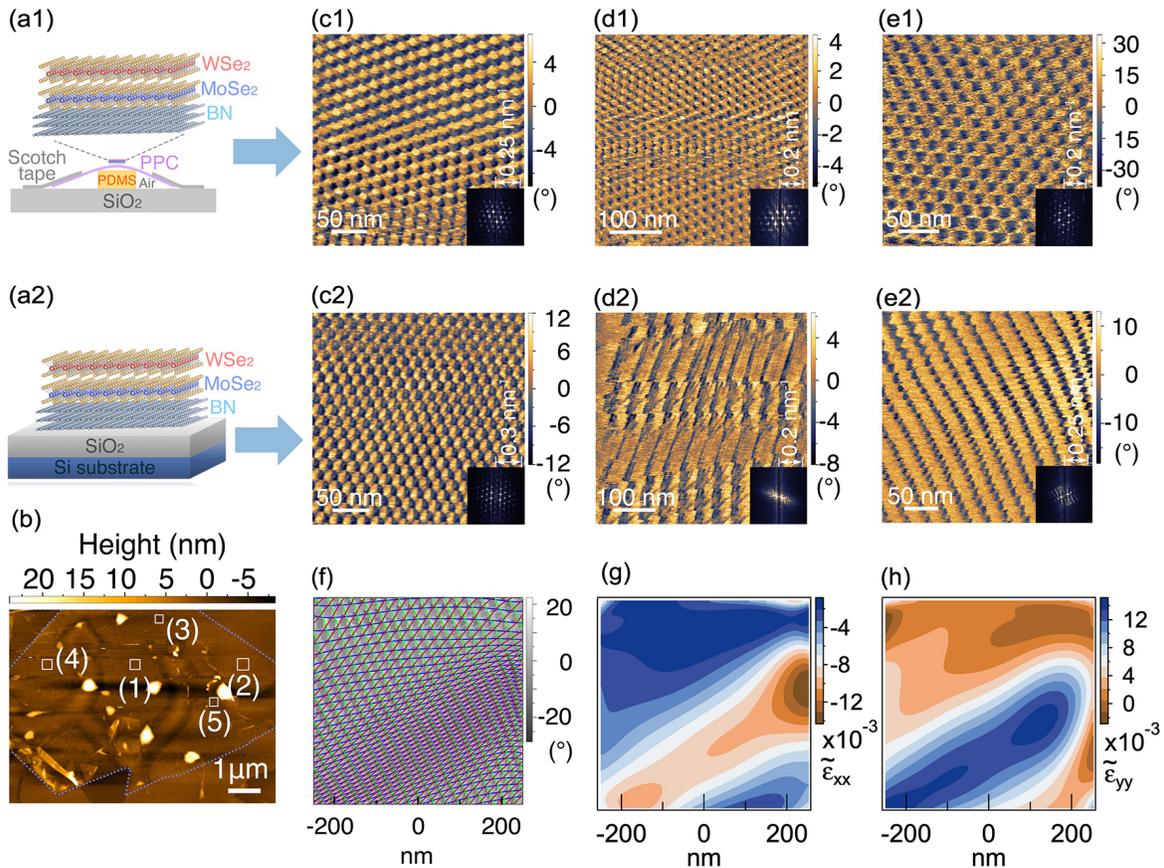

**Fig. 2 | Transformation of Moiré Patterns and Strain Analysis.** (a) Schematics of a BN/WSe$_2$/MoSe$_2$ stack, without top BN capping and with twist angle $\Delta\theta$ = 59.4±0.8°, on (a1) the air-cushioned polypropylene carbonate (PPC) tape; and (a2) the Si/SiO$_2$ surface. (b) AFM image of the WSe$_2$/MoSe$_2$ sample. The dashed line marks the perimeter of the heterobilayer region and the white squares (labeled 1-5) are the spots for PFM imaging before and after transfer form PPC to Si/SiO$_2$. PFM images at three locations are shown before (c1, d1, e1) and after (c2, b2, e2) transfer of the stack from PPC to Si/SiO$_2$. The lower right insets are corresponding FFTs. (f)–(h) Strain field analysis based on a continuum mechanic model for an extended region of the PFM image shown in (e2): 2D grid pattern (f) obtained from the PFM image and strain tensor element maps along (g) the x-axis ($\tilde{\varepsilon}_{xx}$) and (h) the y-axis ($\tilde{\varepsilon}_{yy}$). Note that PFM and AFM are different imaging modes on the same microscope and the locations for PFM imaging are precisely determined from the AFM image.

To establish how differential strain is responsible for the transformation from the 0D hexagonal moiré to the 1D moiré superlattices, we analyze the strain distribution using a continuum mechanics model, as detailed in Supporting Text-I and **Figs. S19-21**. We choose **Fig. 2-e2** with well-resolved hierarchical 1D moiré patterns to analyze the strain field. **Fig. 2f** plots the grid representation of an extend PFM image in **Fig. 2-e2**, showing the strained moiré superlattices. The resulting magnitudes of the normal strain tensor elements $\tilde{\varepsilon}_{xx}$ and $\tilde{\varepsilon}_{yy}$ along the x- and y- directions are shown in **Figs. 2g** and **h**, respectively. As expected, the two strain fields are complementary to each other. The strain maps confirm the anisotropic nature of the strain field which propagates along a direction ~30° from the x-axis (see Supporting Text-I for details). The peak strain is ±1.4% for this particular 1D moiré pattern.

We analyze the excitonic properties derived from these distinct moiré patterns. The type-I PL shown in **Fig. 1e** is consistent with the spectral features reported by Seyler et al,[8] which has been attributed to the 0D moiré excitons in the hexagonal moiré landscape (**Fig. 1b**). In such a moiré landscape, the excitonic potential traps in each superlattice unit cell are located at high-symmetry points with $C_{3v}$ symmetry[13,20], a pre-requisite for preserving the valley optical selection rule and circular polarized PL. At an excitation density of $n_{ex}$ ~ $1 \times 10^{10}$ cm$^{-2}$, we estimate that the total number of interlayer excitons, $N_{ex}$, in a diffraction-limited focal spot (~$5 \times 10^{-8}$ cm$^2$) is about 500. This finite number of excitons can be loaded into a finite number of quantum-dot-like traps, with local variations due to heterogeneity in strain and electrostatics. As a result, one observes quantum dot like PL peaks. We characterize in detail such type-I PL emission on a MoSe$_2$/WSe$_2$ heterobilayer sample with h-BN encapsulation (top and bottom), with a small twist angle $\Delta\theta$ = 58.4 ± 0.4° and sub-nm scale flatness (see **Fig S4**). The observation of sharp PL peaks with circular polarization are in excellent agreement with the recent report[8].

Note that the relatively thick BN capping layer (~20 nm) provides good electrostatic isolation of the MoSe$_2$/WSe$_2$ heterobilayer, enabling the observation of sharp peaks in type-I PL emission, but it prohibits PFM imaging of the moiré patterns underneath. We fabricated another MoSe$_2$/WSe$_2$ heterobilayer with thin (~2 nm) top BN encapsulation on a SiO$_2$/Si substrate, **Fig.**

**S5a**. PFM image through the top BN cap reveals the hexagonal moiré pattern on the MoSe$_2$/WSe$_2$ heterobilayer, **Fig. S5b**, with a superlattice constant of ~ 5.6 nm, corresponding to a twist angle of $\Delta\theta \sim 3.4°$. PL spectra taken from the same spot as in PFM imaging shows partially resolved sharp type-I emission with circular polarized emission under $\sigma^+$ excitation, **Fig. S5c**.

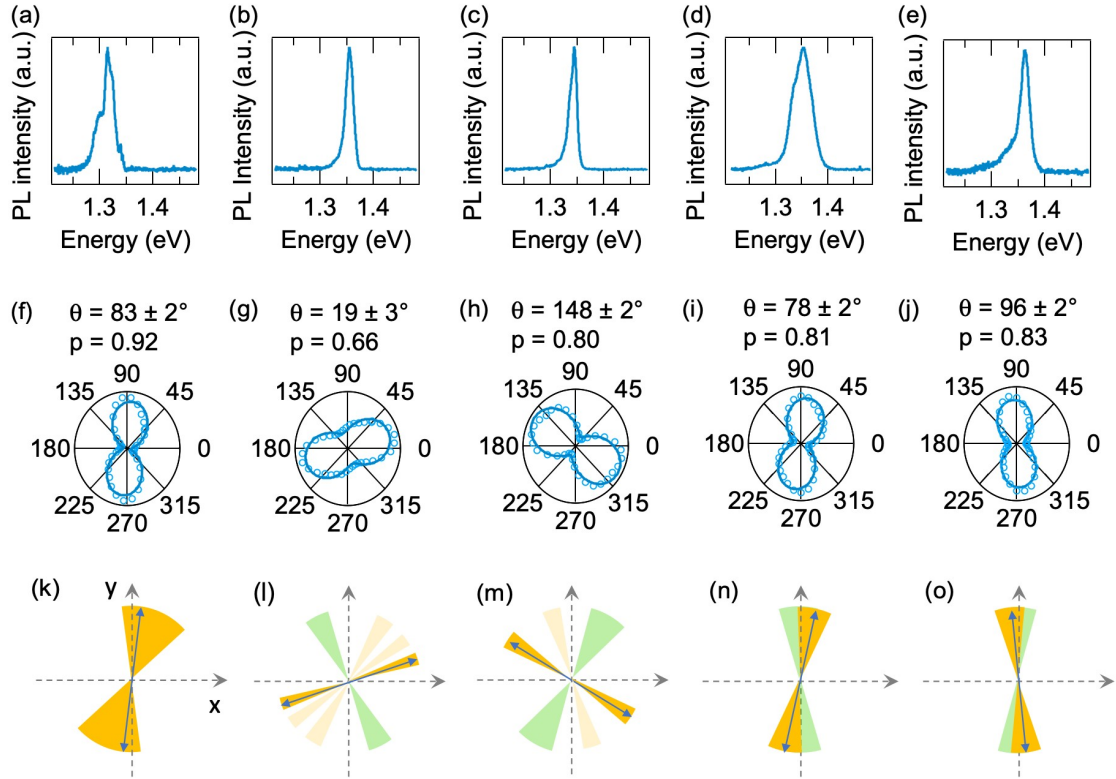

**Fig. 3 | Direct correlation between 1D moiré patterns and linearly polarized PL emission.** (a)-(e) Comparison of the PL optical polarizations with the angle distributions of PFM determined moiré geometrical features; the golden and light golden sectors represent the angular distributions of the primary oval structures in the strained moiré landscapes, the green sectors represent the angular distributions of secondary pseudo-1D stripes formed by the primary structured, while the blue double arrows highlight the PL optical polarization directions. (a) and (b) are derived based on the spots shown in **Fig. 1** (e) and (g), while (c)-(e) are derived based on the spots shown in **Fig. S6**. (f), (g), (h), (i) and (j) Angular dependence of PL emission polarization (blue dots), with fits as blue curves; (f)-(j) are corresponding to the optical polarization illustrated in (a)-(e). θ denotes the phase of the emission polarization, while $p$ denotes the degree of linear polarization, with $p = (I_{max} - I_{min}) / (I_{max} + I_{min})$, where $I_{max}$ ($I_{min}$) is the maximum (minimum) PL emission intensity. (k)-(o) Angular distribution in the 2D FFT of PFM image (colored fans). The double head arrows indicate the long axis of the PL linear polarization in (f)-(j). Excitation density $n_{ex} = 2.5 \times 10^{11}$ cm$^{-2}$ (otherwise same experimental conditions as in **Fig. 1e**).

The type-II PL with a single emission peak, seen in ~80% of the samples with AA or AB stacking on Si/SiO$_2$ substrates, originates from the primary structure in the 1D moiré pattern. We correlate PFM imaging with PL polarization on five representative spots (**Fig. 2-d2, e2**, **Fig. S2-** c, and Fig. S6-b,c) from two different samples. **Figures. 3a-e** show the PL from these five spots with identified 1D moiré patterns. The corresponding polar plots of PL intensity vs linearly polarized detection angle in **Figs. 3f-j** show that all peaks are linearly polarized. The direction of the linear polarization in each case (arrows **Fig. 3k-o**) aligns with the direction of one of the primary structures (golden fans in **Fig. 3k-o**) obtained from 2D-FFT analysis (**Fig. S7**), but not with the direction of secondary 1D moiré structures (green fans in **Fig. 3k-o**).

Note that in **Fig. 3k** (from PFM image in **Fig. 2-d2**) only the direction of the primary 1D moiré structure is resolved. In **Figs. 3l** and **3m** (from PFM image in **Fig. 2-e2** or **Fig. S2-c**) the 2D FFT image analysis gives 2-3 apparent directions (golden fans), but only one of them represents the real primary 1D direction which aligns with the linear PL polarization direction. In **Figs. 3n** and **3o** (from PFM image in **Fig. S6-b, c**), the directions of the primary 1D structures are not clearly resolved from that of the secondary moiré structures. Linearly polarized PL emission is well known from semiconductor nanowires and carbon nanotubes[21,22]. Thus, we take the linear PL emission and its spatial correlation with the 1D moiré morphology as evidence for the 1D moiré potential.

We analyze the type-II PL emission in detail using a MoSe$_2$/WSe$_2$ heterobilayer sample sandwiched by h-BN (**Fig. S8**), which is known to reduce electrostatic heterogeneity, decrease PL peak width, and increase PL intensity[23]. **Fig. S9** summarizes fitting results to the PL from multiple spots on the PL map. Importantly, the direction of linear polarization, which we correlate with the direction of the primary structure in 1D moiré strips, evolves from spot to spot, suggesting the spatial evolution of strain fields in the sample. On the other hand, the PL intensity is independent of the polarization of excitation light, at the intralayer exciton energy, as shown by the isotropic dependence (orange dots in **Fig. S9-e**). Thus, we infer that the 1D moiré pattern has negligible effect on the intralayer excitons, where moiré potential is expected to be shallower than that for interlayer excitons[1]. The hetero-strain field can vary locally depending on details of fabrications and sample condition, such as the presence of bubbles in the heterobilayer or between the heterobilayer and BN encapsulation layers. Both 0D and 1D moiré traps can co-exist at different regions on the same sample. **Figure S10** shows both type-I and type-II PL at different locations on a single MoSe$_2$/WSe$_2$ heterobilayer sample with $\Delta\theta = 2.6 \pm 0.5°$. In contrast to the linearly polarized type-II PL, the circularly polarized type-I PL in **Fig. S4** is close-to-isotropic in azimuthal angle-resolved polarization distribution (**Fig. S11**).

Note that the widths of type-II PL from fully h-BN encapsulated samples were FWHM = 8±2 meV (see lower spectrum in **Fig. 1e**, lower two spectra in **Fig. S9c**, the four spectra from different spots in **Fig. S10e**, and the spectra at low excitation densities in Fig. 1a in ref. 12). For comparison, the PL peak width in the absence of top h-BN excapsulation doubles to FWHM = 16±4 meV, but the asymmetric peak shape is retained, as shown in **Fig. 3b** and **3c**. This broadening can be attributed to dielectric heterogeneity[24], resulting, e.g., from the adsorption of molecules on the MoSe$_2$/WSe$_2$ sample surface under cryogenic measurements. On some spots, the broadened PL peak can be attribtued to the sum of 2 peaks, as in **Fig. 3a**, **3d**, and **3e**. This likely results from the presence of two different dielectric environment or two domains of stained moiré patterns with ~ 1 μm diameter spot of the probe region.

The formation of 1D moiré excitons on a strained moiré landscape explains key properties of type-II PL. The loss of valley-contrasting properties, i.e. the absence of circular polarized emission, is attributed to the breaking of $C_3$ rotational symmetry by the 1D moiré trap. Thus, the electron-hole exchange can hybridize the two valley configurations of excitons and lead to a linearly polarized state[25]. We analyze the contrasting PL features between type-I and type-II at a qualitative level with a 2D harmonic trap model (Supporting Text-II and **Figs. S22a & b**). Since the process of forming interlayer excitons from higher energy intralayer excitons is always accompanied by excess energy, which amounts to ~300 meV for the MoSe$_2$/WSe$_2$ heterobilayer[12,15,16], radiative recombination may occur from excitonic levels higher than the lowest energy state. A 0D moiré potential hosts very few bound states and the lowest exciton state likely dominates PL emission, leading to quantum emitter-like sharp PL peaks. Compared to the 0D moiré trap, a 1D moiré trap is characterized by higher DOS with densely spaced excitonic levels, **Fig. S22c**, radiative recombination from multiple levels near the bottom of the potential well may occur, leading to both broadening in PL peak and increase in total PL intensity. Moreover, additional enhancement of PL may result from delocalization when the primary 1D moiré traps are closely packed and merge into the secondary 1D moiré patterns, as suggested by PFM images. In all MoSe$_2$/WSe$_2$ heterobilayer samples investigated here, the PL intensity vary greatly from location to location on each sample, as shown in confocal PL image maps in Fig. S4c, Fig. S9b, and Fig. S10d. Such variation in PL intensity may be attributed to heterogeneity in strain fields that result in variations in local DOS as well as in extents of localization/delocalization.

Since both 0D and 1D moiré excitons result from relatively shallow local potential traps on a 2D landscape, we expect interlayer excitons to also form outside these traps at sufficiently high excitation densities due to mutual repulsion among these excitons with permanent electric dipoles[1]. These are confirmed for the 1D moiré excitons in **Fig. 4**. The first evidence comes from the reduction of linear polarization. As shown in **Fig. 4a**, when the excitation density increases from $n_{ex} = 6.1 \times 10^9$ to $7.3 \times 10^{12}$ cm$^{-2}$, the purity of linear polarization, $p = (I_{max} - I_{min}) / (I_{max} + I_{min})$, decreases from 0.89 to 0.63. The second evidence comes from the evolution in peak shapes. As shown in **Fig. 4b**, with increasing $n_{ex}$, the single PL peak attributed to the 1D moiré exciton evolves into a multi-peak shape, with growth of shoulder features on both lower and higher energy sides. This evolution is particularly evident above the Mott density ($n_{Mott} \sim 3 \times 10^{12}$)[12].

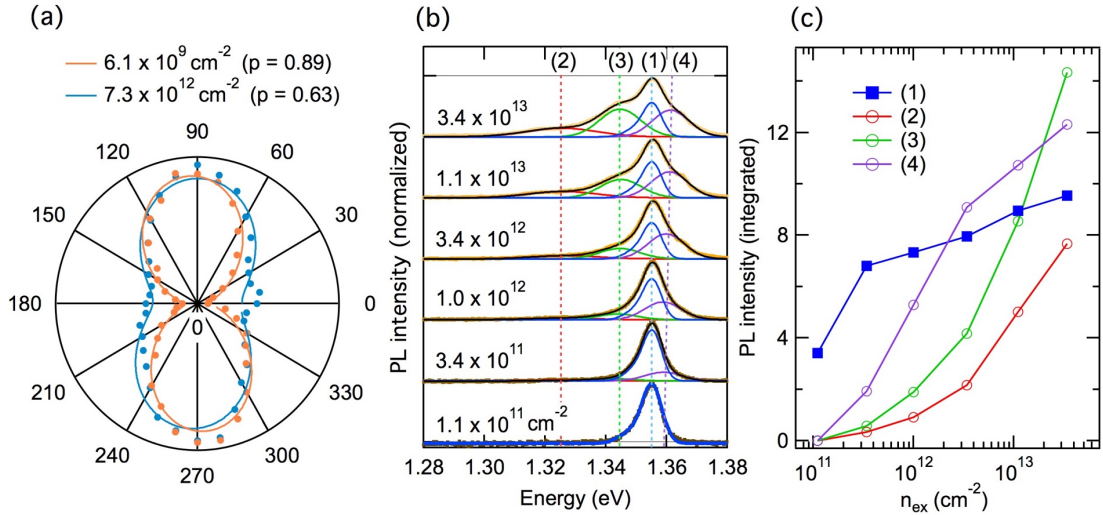

**Fig. 4 | Evolution of PL peak shape with exciton density for the 1D moiré excitons.** The data are obtained on the same sample as used in Fig. 2. (a) Linear-polarization distributions of PL emission at two excitation densities; the degree of polarization, $p = 0.89$ and $0.63$, at $n_{ex} = 6.1 \times 10^9$ cm$^{-2}$ and $7.3 \times 10^{12}$ cm$^{-2}$, respectively. (b) Intensity normalized PL spectra (dots) acquired with varying excitation densities $n_{ex} = 1.1 \times 10^{11}$ cm$^{-2}$ - $3.4 \times 10^{13}$ cm$^{-2}$. The colored curves are deconvolutions detailed in the text. The dashed lines mark the peak positions (1-4) of the four deconvoluted peaks. (c) Integrated intensities of the deconvoluted PL peaks shown in (b) as a function of excitation density. Sample temperature = 4 K.

The solid curves in **Fig. 4b** show approximate deconvolutions of the spectra into the main asymmetric PL peak (1) and three Gaussian peaks, with positions shown by the dashed lines. Peaks (2) and (3) are on the lower energy side of the main peak, and peak (4) is on the higher energy side. Similarly, we observe the evolution of the initially quantum-emitter like type-I moiré excitons into broader multiple PL peak-shapes as excitation density is increased above $n_{Mott}$, as shown in **Figs. S12**, **S13**. These results suggest the radiative recombination from increased population of interlayer excitons outside the 0D or 1D moiré traps at sufficiently high excitation density, as is demonstrated also for both type-I and type-II PL from the same sample (**Fig. S14**). The exact origin of multiple PL peak shapes for interlayer excitons in the MoSe$_2$/WSe$_2$ heterobilayer at high excitation densities is not well understood at the present. A recent report by Tran et al.[9] attributed similar multiple peaks in the WSe$_2$/MoSe$_2$ heterobilayer to quantized levels within the 0D moiré trap. Here we observe the multiple PL peaks in most WSe$_2$/MoSe$_2$ heterobilayer samples only at high excitation densities, regardless of the initial nature of the moiré excitons (type-I or type-II). This suggests that the four-peak structure originates from radiative recombination of free interlayer excitons outside the moiré traps.

In both MoSe$_2$/WSe$_2$ and MoS$_2$/WS$_2$ heterobilayers, density functional theory (DFT) calculations suggest that the conduction band minima are at the $Q$ point, not the $K$ point[26,27]. Thus, the two emission peaks on the lower energy side of the main PL peak may be attributed to the spin-orbital split conduction bands at $Q$ involving the momentum indirect excitons $Q_C$-$K_V$, where the subscript denote the conduction ($C$) and valence ($V$) bands, respectively. Previous pump-probe reflectance and time-resolved PL measurements on WSe$_2$/MoSe$_2$ heterobilayers have revealed efficient scattering from $K$ to $Q$ valleys in the Brillouin zone[12]. In the related system of WS$_2$/MoS$_2$, a time- and angle-resolved photoemission spectroscopy (TR-ARPES) study provided direct evidence for photo-excited electrons in the $Q$ valleys that are of strongly mixed character of both TMDC monolayers[28]. Similarly, the emission peak on the high energy side may come from the momentum indirect $K^*_C$-$K_V$' involving the upper spin-orbital split conduction band ($^*$). Radiative emission of these momentum-indirect excitons can be facilitated at high density through the pair-annihilation, with momentum conservation satisfied, as detailed in **Fig. S15**. Similar to the mechanism proposed by Rivera et al.[29] (see Fig. S12D in ref. [29]), a pair of inter-valley dark excitons, $K^*_C$-$K_V$' (blue arrow) and $K^*_C$'-$K_V$ (blue dashed arrow) can virtually swap the electron-hole pairing and thus emit two photons through the intermediate state of two bright excitons (peak 4). Similarly, for a pair of dark $Q_C$-$K_V$ and $Q_C$'-$K_V$' excitons, the pairwise $Q$ to $K$ and $Q$' to $K$' scattering of their electrons into two virtual $K_C$-$K_V$ and $K_C$'-$K_V$' bright excitons can provide the radiative emission channel (peaks 2 and 3). Radiative recombination from the pair annihilation of dark excitons is expected at higher exciton density, in comparison with the normal PL emission from the bright $K_C$-$K_V$ exciton. Supporting this, we note that the multiple peaks (2-4) attributed to momentum-indirect excitons are vanishingly small at low excitation densities ($\leq 1\times10^{11}$ cm$^{-2}$) but grow much faster with $n_{ex}$ than peak (1), **Fig. 4c**.

The discovery of the 1D moiré potential landscape from uniaxial strain and the corresponding 1D moiré excitons in TMDC heterobilayers has two implications in the growing field of moiré physics. On the one hand, given the susceptibility of 2D crystals to strain within samples consisting of multilayers created by mechanical transfer stacking, one must consider the presence of strain fields in understanding the moiré potential and its heterogeneity in determining physical properties. On the other hand, this very sensitivity to strain field provides a new method to control the moiré potential landscape in hetorobilayers, leading to optical and electronic responses on-demand. The 1D moiré potential is particularly attractive for strongly correlated and anisotropic charge transport[4,30]. In comparison to the large strain required to bring in significant changes in monolayers of graphene[31–33] and TMDCs[34–36], moiré strain engineering can be achieved with a modest differential strain on the order of the lattice mismatch[4]. Combined with the twist angle, active or passive control of strain fields thus open the door to greater opportunities of artificially generating band structures[30] and topological mosaics[4] at 2D material interfaces.

**MATERIALS and METHODS**

**TMDC monolayers.** Monolayers of WSe$_2$ and MoSe$_2$ were mechanically exfoliated from bulk crystals grown by the self-flux method. These monolayers possessed low defect densities (<10$^{11}$ cm$^{-2}$)[37]. Flakes of h-BN with thickness 5 – 35 nm and with atomically flat surfaces were obtained by mechanical exfoliation. The flakes (WSe$_2$, MoSe$_2$, and h-BN) were characterized by atomic force microscopy (AFM) and Raman spectroscopy.

**Determination of TMD zigzag crystal orientation via SHG.** The crystal orientations of WSe$_2$ and MoSe$_2$ monolayers were determined by second harmonic generation (SHG) measurement on an inverted optical microscope (Olympus IX73). Linearly polarized femtosecond laser light (Spectrum Physics Tsunami, 80 MHz, 800 nm, 80 fs) was focused onto a monolayer with a 100x, NA 0.80 objective (Olympus LMPLFLN100X). The reflected SHG signal at 400 nm was collected by the same objective, filtered by a short-pass dichroic mirror, short-pass and band-pass filters, and a Glan-Taylor linear polarizer, detected by a photomultiplier tube (Hamamatsu R4220P), and recorded by a photon counter (BK PRECISION 1823A 2.4GHz Universal Frequency Counter).

We obtain the azimuthal angular (θ) distribution of SHG signal by rotating the laser polarization and the SHG signal (via a half waveplate) with fixed sample orientation. Due to the $D_{3h}$ symmetry, the non-vanishing tensor elements of the second order susceptibility of WSe$_2$ and MoSe$_2$ monolayers are $\chi^{(2)}_{yyy} = -\chi^{(2)}_{yxx} = -\chi^{(2)}_{xxy} = -\chi^{(2)}_{xyx}$ where the $x$ axis is defined as the zigzag direction (*1*). When we simultaneously rotated the fundamental and SHG signals, the SHG intensity showed six-fold symmetry: $I_\perp \propto cos^2(3\theta)$ and $I_\parallel \propto sin^2(3\theta)$, where $\theta$ is the angle between the laser polarization and the zigzag direction. We use triangular flakes of monolayer WS$_2$ (6Carbon) or MoS$_2$ (2DLayer), where zigzag directions are the same as crystal edges, both grown from chemical vapor deposition (CVD), to calibrate the SHG setup.

**Preparation of 2D WSe$_2$/MoSe$_2$ heterostructure samples.** The 2D WSe$_2$/MoSe$_2$ heterobilayer samples were prepared from the polymer-free van der Waals assembly technique[11]. A transparent polydimethylsiloxane (PDMS) stamp coated with a thin layer of polypropylene carbonate (PPC) was used to pick up a thin layer of exfoliated h-BN. This h-BN was then used to pick up the first TMDC monolayer. The second TMDC monolayer was aligned to and picked up by the first monolayer on a rotation stage. After picking up a second BN flake as the bottom encapsulation layer, we transferred the entire structure on to a clean silicon wafer (with 285 nm thermal oxide layer for enhanced optical contrast) at elevated temperatures (90-130 °C). The residual PPC was washed away by acetone to give a clean h-BN/MoSe$_2$/WSe$_2$/h-BN heterobilayer on the Si/SiO$_2$ substrate. The samples were then thermal annealed in an ultrahigh vacuum chamber ($10^{-8}$-$10^{-9}$ Torr): the sample temperature was raised from room temperature to 523K slowly over two hours, and kept at this temperature (523K) for an additional three hours; then the sample was cooled down to 173K over 30 min, after which the sample was eventually raise up to room temperature over 12 hours.

**Polarization-resolved confocal microscopic measurements.** PL imaging was performed on a home-built scanning confocal microscope system (**Fig. S16**) based on a liquid-helium recirculating optical cryostat (Montana Instruments Fusion/X-Plane) with a 100x, NA 0.75 objective (Zeiss LD EC Epiplan-Neofluar 100x/0.75 HD DIC M27). The temperature of the sample stage could be varied between 3.8 K and 350 K. In all experiments presented in this study, the TMDC heterobilayer and monolayers samples were at 4 K in a vacuum (<$10^{-6}$ torr) environment, unless otherwise noted. A Galvo-Galvo scanner (Thorlabs, GVS012/M) was used for mapping the PL signal emitted from the sample plane. Polarizers, λ/2 waveplate, and λ/4 waveplate were used for circular/linear polarization-resolved experiments (see Fig. S11 for the detailed optical setup). The incident laser beam (Coherent Rega, 750 nm, ~ 150 fs, 76 MHz) was focused by the objective to a diffraction limited spot on the sample. The excitation power was measured by a calibrated power meter (Ophir StarLite) with broad dynamic range. The PL light was collected by the same objective, spatially and spectrally filtered, dispersed by a grating, and detected by an InGaAs photodiode array (Princeton Instruments PyLoN-IR). The wavelength was calibrated by neon-argon and mercury atomic emission sources (Princeton Instruments IntelliCal). In all PL measurements, we found no laser heating of the sample under cryogenic cooling. The PL spectra are completely reproducible following repeated measurements at the same spot on each sample (e.g., **Fig. S17**).

**Piezoresponse force microscopy measurements.** All PFM imaging experiments were performed on a commercially available Bruker Dimension Icon AFM with a Nanoscope IIIa controller. We used ASYELEC-01 Ti/Ir coated silicon probes with a force constant of ~3 N/m from Oxford Instruments Asylum Research. The amplitudes of AC bias were <1 V and the single frequency excitation was at the lateral cantilever-sample resonance in the range 750-850 kHz. Most PFM imaging experiments were carried out on $WSe_2/MoSe_2$ heterobilayer samples without the top BN encapsulation layer, with the sample stack either on PPC/PDMS or transferred onto $SiO_2$/Si. Similar PFM images have been obtained for $WSe_2/MoSe_2$ heterobilayer samples with thin BN capping layers (**Fig. S5**), albeit at reduced resolution as compared to the exposed $WSe_2/MoSe_2$ heterobilayer. Note that thermal annealing procedure used on some of the heterobilayer samples did not result in changes to PFM images for either 0D or 1D moiré patterns (see, **Fig. S18**).

**Acknowledgement.** The PFM imaging experiments and PL measurements were supported by the Center for Programmable Quantum Materials, an Energy Frontier Research Center funded by the US Department of Energy, grant DE-SC0019443. Sample preparation and SHG characterization were supported by the Center for Precision Assembly of Superstratic and Superatomic Solids, a Materials Science and Engineering Research Center (MRSEC) through NSF grant DMR-1420643. Building of the confocal PL spectrometer was supported in part by the Office of Naval Research under award no. N00014-16-1-2921.

**References:**

1. Yao, W., Xu, X., Liu, G.-B., Tang, J. & Yu, H. Moiré excitons: From programmable quantum emitter arrays to spin-orbit–coupled artificial lattices. *Sci. Adv.* **3,** e1701696 (2017).
2. Cao, Y. *et al.* Unconventional superconductivity in magic-angle graphene superlattices. *Nature* **556,** 43–50 (2018).
3. Cao, Y. *et al.* Correlated insulator behaviour at half-filling in magic-angle graphene superlattices. *Nature* **556,** 80–84 (2018).
4. Tong, Q. *et al.* Topological mosaics in moiré superlattices of van der Waals heterobilayers. *Nat. Phys.* **13,** 356–362 (2017).
5. Fogler, M. M., Butov, L. V & Novoselov, K. S. High-temperature superfluidity with indirect excitons in van der Waals heterostructures. *Nat. Commun.* **5,** (2014).
6. Jin, C. *et al.* Observation of moiré excitons in WSe2/WS2 heterostructure superlattices. *Nature* **567,** 76–80 (2019).
7. Alexeev, E. M. *et al.* Resonantly hybridized excitons in moiré superlattices in van der Waals heterostructures. *Nature* **567,** 81–86 (2019).
8. Seyler, K. L. *et al.* Signatures of moiré-trapped valley excitons in MoSe2/WSe2 heterobilayers. *Nature* **567,** 66–70 (2019).

9. Tran, K. *et al.* Evidence for moiré excitons in van der Waals heterostructures. *Nature* **567,** 71–75 (2019).
10. Wu, F., Lovorn, T. & Macdonald, A. H. Topological Exciton Bands in Moiré Heterojunctions. *Phys. Rev. Lett.* **118,** 147401 (2017).
11. Wang, L. *et al.* One-dimensional electrical contact to a two-dimensional material. *Science* **342,** 614–7 (2013).
12. Wang, J. *et al.* Optical generation of high carrier densities in 2D semiconductor heterobilayers. *Sci. Adv.* **5,** eaax0145 (2019).
13. Rivera, P. *et al.* Interlayer valley excitons in heterobilayers of transition metal dichalcogenides. *Nat. Nanotechnol.* **13,** 1004–1015 (2018).
14. Li, Y. *et al.* Probing symmetry properties of few-layer MoS2 and h-BN by optical second-harmonic generation. *Nano Lett.* **13,** 3329–3333 (2013).
15. Rivera, P. *et al.* Observation of long-lived interlayer excitons in monolayer $MoSe_2$-$WSe_2$ heterostructures. *Nat. Commun.* **6,** 6242 (2015).
16. Jauregui, L. A. *et al.* Electrical control of interlayer exciton dynamics in atomically thin heterostructures. *Science* **366,** 870–875 (2019).
17. Wu, W. *et al.* Piezoelectricity of single-atomic-layer MoS 2 for energy conversion and piezotronics. *Nature* **514,** 470–474 (2014).
18. Lee, J. H. *et al.* Reliable Piezoelectricity in Bilayer WSe 2 for Piezoelectric Nanogenerators. *Adv. Mater.* **29,** 1–7 (2017).
19. Mcgilly, L. J. *et al.* Seeing moiré superlattices. *To be Publ.* 1–17
20. Liu, G. *et al.* Electronic structures and theoretical modelling of two-dimensional group-VIB transition metal dichalcogenides. *Chem. Soc. Rev.* **44,** 2643–2663 (2014).
21. Wang, J., Gudiksen, M. S., Duan, X., Cui, Y. & Lieber, C. M. Highly polarized photoluminescence and photodetection from single indium phosphide nanowires. *Science* **293,** 1455–1457 (2001).
22. Lefebvre, J., Fraser, J. M., Finnie, P. & Homma, Y. Photoluminescence from an individual single-walled carbon nanotube. *Phys. Rev. B* **69,** 75403 (2004).
23. Ajayi, O. *et al.* Approaching the Intrinsic Photoluminescence Linewidth in Transition Metal Dichalcogenide Monolayers. *2D Mater.* 10.1088/2053-1583/aa6aa1 (2017). at <http://iopscience.iop.org/10.1088/2053-1583/aa6aa1>

24. Raja, A. *et al.* Dielectric disorder in two-dimensional materials. *Nat. Nanotechnol.* **14,** 832–837 (2019).

25. Yu, H., Liu, G.-B., Gong, P., Xu, X. & Yao, W. Dirac cones and Dirac saddle points of bright excitons in monolayer transition metal dichalcogenides. *Nat. Commun.* **5,** 3876 (2014).

26. Gillen, R. & Maultzsch, J. Interlayer excitons in MoSe2/WSe2 heterostructures from first principles. *Phys. Rev. B* **97,** 1–7 (2018).

27. Okada, M. *et al.* Direct and Indirect Interlayer Excitons in a van der Waals Heterostructure of hBN/WS2/MoS2/hBN. *ACS Nano* **12,** 2498–2505 (2018).

28. Liu, F. & Zhu, X. Direct determination of momentum resolved electron transfer in the photo-excited MoS2/WS2 van der Waals heterobilayer by TR-ARPES. *arXiv Prepr.* **1909,** 7759 (2019).

29. Rivera, P. *et al.* Valley-polarized exciton dynamics in a 2D semiconductor heterostructure. *Science* **351,** 688–691 (2016).

30. Bi, Z., Yuan, N. F. Q. & Fu, L. Designing flat bands by strain. *Phys. Rev. B* **100,** 35448 (2019).

31. Ni, Z. H. *et al.* Uniaxial strain on graphene: Raman spectroscopy study and band-gap opening. *ACS Nano* **2,** 2301–2305 (2008).

32. Gui, G., Li, J. & Zhong, J. Band structure engineering of graphene by strain: first-principles calculations. *Phys. Rev. B* **78,** 75435 (2008).

33. Guinea, F., Katsnelson, M. I. & Geim, a. K. Energy gaps, topological insulator state and zero-field quantum Hall effect in graphene by strain engineering. *Nat. Phys.* **6,** 30–33 (2009).

34. Castellanos-Gomez, A. *et al.* Local strain engineering in atomically thin MoS2. *Nano Lett.* **13,** 5361–5366 (2013).

35. He, K., Poole, C., Mak, K. F. & Shan, J. Experimental demonstration of continuous electronic structure tuning via strain in atomically thin MoS2. *Nano Lett.* **13,** 2931–2936 (2013).

36. Kou, L., Frauenheim, T. & Chen, C. Nanoscale multilayer transition-metal dichalcogenide heterostructures: Band gap modulation by interfacial strain and spontaneous polarization. *J. Phys. Chem. Lett.* **4,** 1730–1736 (2013).

37. Edelberg, D. *et al.* Approaching the Intrinsic Limit in Transition Metal Diselenides via Point Defect Control. *Nano Lett.* **19,** 4371–4379 (2019).